\documentstyle[12pt]{article}

\def\pd{\partial}
\def\a{\alpha}
\def\b{\beta}

\def\eps{\epsilon}

\def\e{{\rm e}}
\def\z{{\bar z}}
\def\w{{\bar w}}
\def\half{\frac{1}{2}}
\def\fr{\frac}
\def\T{{\hat T}}
\def\Tp{{\hat T}^+}
\def\Tm{{\hat T}^-}
\def\S{{\hat S}}
\def\pp{\prime}
\def\bb{\begin{equation}}
\def\ee{\end{equation}}
\def\bba{\begin{eqnarray}}
\def\eea{\end{eqnarray}}

\begin{document}

\begin{titlepage}

\begin{tabbing}
   qqqqqqqqqqqqqqqqqqqqqqqqqqqqqqqqqqqqqqqqqqqqqq
   \= qqqqqqqqqqqqq  \kill
         \>  {\sc KEK-TH-449} \\
         \>   hep-th/9509025 \\
         \>  {\sc September 1995}
\end{tabbing}
\vspace{5mm}

\begin{center}
{\Large {\bf Ward Identities of $W_{\infty}$ Symmetry \break
and Higher Genus Amplitudes \break
in 2D String Theory}}
\end{center}

\vspace{1cm}

\centering{\sc Ken-ji HAMADA}\footnote{E-mail address : hamada@theory.kek.jp}

\vspace{7mm}

\begin{center}
{\it National Laboratory for High Energy Physics (KEK),} \\
{\it Tsukuba, Ibaraki 305, Japan}
\end{center}

\vspace{3mm}

\begin{center}
({\it Revised version published in Nucl. Phys. B})
\end{center}

\vspace{3mm}

\begin{abstract}
The Ward identities of the $W_{\infty}$ symmetry in two dimensional
string theory in the tachyon background are studied in the continuum
approach. We consider amplitudes different from 2D string ones
by the external leg factor and derive the recursion relations among
them. The recursion relations have non-linear terms which give
relations among the amplitudes defined on different genus.
The solutions agree with the matrix model results even in higher genus.
We also discuss differences of roles of the external leg factor between
the $c_M = 1$ model and the $c_M <1$ model.
\end{abstract}
\end{titlepage}

\section{Introduction and Summary}
\indent

   Many interesting issues in string theory such as dynamically
compactification, black hole physics, etc,  require a non-perturbative
formulation. Such a formulation is not now available in higher
dimensional string theories. In two or fewer spacetime dimensions,
however, the string theory becomes solvable~\cite{k}
due to the presence of an infinite number of
$W_{\infty}$ currents~\cite{w,k2,h,fkn2,im,hop,dmw},
which gives a possibility of studying the non-perturbative formulation
of string theory. Furthermore 2D string theory itself is interesting
in spacetime physics. It gives the 2D dilaton gravity with a massless
matter called ``tachyon'' as the effective theory~\cite{np}.
Thus 2D string theory is also attractive as an alternative approach
to studying 2D quantum dilaton gravity~\cite{h2}.

  There are several formulations of two dimensional string theory.
The matrix model (see reviews~\cite{k})
is generally believed to describe the 2D string
theory, which is in principle defined non-perturbatively.
The continuum theory~\cite{dk,s,p}
is defined using the standard quantization
method of the string perturbation theory.
The topological description~\cite{l} of 2D string theory is formulated
in~\cite{mv,hop}.
To understand the non-perturbative formulation of string theory
it is important to clarify the relations between these methods.
In the present work we investigate the continuum method of 2D string
theory. We consider the Ward identities of the $W_{\infty}$ symmetry
for amplitudes in the tachyon background and give the recursion
relations which connect amplitudes on different genus.
We then study the correspondence between the matrix model and the
continuum theory genus by genus.

  String theory in 2D target space-time is described by a
combination of the Liouville and the $c_M =1$ matter fields. The
Liouville and the matter fields are identified with the space and
the time coordinates in the target space. In the next section we
define the $S$-matrix of the massless tachyons in the tachyon
background. Here we introduce the $\S$-matrix, which is different
from the $S$-matrix by the external leg factor. We will see that the
$\S$-matrix is equivalent to the amplitude of the $c_M =1$ matrix model
even in higher genus. In Sect.3 we set up the Ward identities of
the $W_{\infty}$ symmetry~\footnote{
The Ward identities of $W_{\infty}$ symmetry for $c_M =1$ theory were
first discussed in ref.~\cite{k2}. They considered the {\it linear}
Ward identities, which, however, do not give correct answers except
for special cases of amplitudes.
In general we need the non-linear terms as discussed in this paper.}.
We give the boundary formulas contributing to the identities,
which are obtained by taking the
$c_M \rightarrow 1$ limit of the results computed in the previous work
for the $c_M < 1$ model~\cite{h}.
In Sect.4 we write out several Ward identities
using the contributions calculated in Sect.3  and solve them.
We first consider sphere amplitudes. In this case we can give the
procedure to derive all types of sphere amplitudes. We first give the
derivation of the $1 \rightarrow N$ amplitudes and then
the $2 \rightarrow N$ ones. Using this  ordering we can recursively
obtain all further types of sphere amplitudes.
We here explicitely calculate the $1 \rightarrow N$,
the $2 \rightarrow 2$ and the $2 \rightarrow 3$ amplitudes for all
kinematical regions, which agree with the results calculated in the
matrix model~\cite{k}
and also the topological 2D string~\cite{hop}.
For higher genus cases we need further discussions. For the current
$W_{-n,-m}$ with $n,m \leq 2$, the identities are closed within the
contributions calculated in Sect.3. But, for the general $W_{\infty}$
currents, extra boundary contributions are needed. Their direct
calculations are very difficult. Here, we guess the formula
from a simple argument and check that the identities are indeed
closed. We also give the procedure to calculate  higher genus
amplitudes recurrently. Thus we confirm that the solutions of the
identities agree with the matrix model results~\cite{k,mpr}
up to genus three.
In the last part of this section we also calculate the
partition function of 2D string theory. In Sect.5 we discuss differences
of structures, especially roles of the external leg factor, between
the $c_M =1$ model and the $c_M <1$ model.

\section{Scattering Amplitudes of Tachyons}
\indent

   Two dimensional string theory is defined by the
action~\footnote{Note that the normalization of the fields differs from
that in ref.~\cite{h} by $\sqrt 2$.}
\bb
    I_0 = \fr{1}{4\pi}\int d^2 z \mbox{$\sqrt{\hat g}$}
            ({\hat g}^{\a \b} \pd_{\a} \phi \pd_{\b} \phi
               + {\hat g}^{\a \b} \pd_{\a} X \pd_{\b} X
               + 2 {\hat R} \phi ) ~,
\ee
where $\phi$ is the Liouville field and $X$ is the $c_M =1$ matter
field. In spacetime physics $\phi$ is identified with the space
coordinate and $X$ is the (Euclidean) time.

   A physical state with continuous momentum can only be the massless
scalar called  ``tachyon'' in the string terminology.
In two dimensions the tachyon mode becomes massless.
Besides that there are an infinite number of the physical states
that exist only at discrete momenta, called the discrete
states~\cite{bmp}.
They are discussed in the next section.

  The tachyon vertex operator with momentum $k ~(>0)$ is
given by
\bb
     T^{\pm}_k =  \fr{1}{\pi} \int d^2 z
         \e^{(2-k)\phi(z,\z) \pm ikX(z,\z)} ~,
\ee
where $\pm$ denotes the chirality.
The selection of $k>0$ is called the Seiberg condition~\cite{s}.
We will postulate below that the amplitude including the anti-Seiberg
$(k<0)$ states vanishes.
Henceforth we introduce the normalized operator
\bb
     \T^{\pm}_k = \Lambda(k) T^{\pm}_k  ~,
               \qquad \Lambda(k) =  \fr{\Gamma(k)}{\Gamma(-k)} ~.
\ee

  Let us consider the action in the tachyon background
\bb
          I= I_0 + \mu_B T_0 ~,
\ee
where $T_0 = \lim_{\eps \rightarrow 0} \T^{\pm}_{\eps}$.
The bare tachyon background $\mu_B$ is divergent,
which is renormalized as follows:
\bb
       \mu_B ~ \eps \rightarrow \mu ~.
\ee
This ensures the non-decoupling of $\mu_B T_0$.

  The $S$-matrix of tachyons in the tachyon background is defined by
\bb
   S^{(g)}_{k_1,\cdots, k_N \rightarrow p_1, \cdots, p_M}
    = \prod^N_{i=1} \Lambda^{-1}(k_i) \prod^M_{j=1} \Lambda^{-1}(p_j)
       ~\S^{(g)}_{k_1,\cdots, k_N \rightarrow p_1, \cdots, p_M} ~,
\ee
where the $\S$-matrix is
\bba
   &&  \S^{(g)}_{k_1,\cdots, k_N \rightarrow p_1, \cdots, p_M}
          = < \prod_{i=1}^N \Tp_{k_i} \prod_{j=1}^M \Tm_{p_j}>_g  \\
   &&  =  \biggl( -\fr{\lambda}{2} \biggr)^{-\chi/2}
            ~\delta \biggl(  \sum^N_{i=1} k_i - \sum^M_{j=1} p_j \biggr)
                  ~\mu_B^s \fr{\Gamma(-s)}{2}
               < \prod_{i=1}^N \Tp_{k_i} \prod_{j=1}^M \Tm_{p_j}~
                   (T_0)^s  >_g^{(free)} ~,
                      \nonumber
\eea
The superscript $free$ denotes the free field representation.
The $\delta$-function and $\mu_B^s \fr{\Gamma(-s)}{2}$ come from the
zero-mode integrals of $X$ and $\phi$ respectively~\cite{gl}.
$g$ is the genus, $\chi =2-2g$ and $s$ is given by
\bb
        s= \sum_{i=1}^N k_i + \chi -N-M ~.
\ee
Thus we introduce an $\S$-matrix different from the $S$-matrix of
two dimensional string theory by the external leg factor $\Lambda^{-1}$.
We will see that the $\S$-matrix is equivalent to the amplitudes
of the $c_M=1$ matrix model even in higher genus.

   Finally we note that the $\S$-matrix is invariant under the
interchange of chiralities,
\bb
       < \prod_{i=1}^N \Tp_{k_i} \prod_{j=1}^M \Tm_{p_j}>_g
        = < \prod_{i=1}^N \Tm_{k_i} \prod_{j=1}^M \Tp_{p_j}>_g
\ee
or $\S^{(g)}_{k_1,\cdots, k_N \rightarrow p_1, \cdots, p_M} =
\S^{(g)}_{p_1, \cdots, p_M \rightarrow k_1,\cdots, k_N}$.

\section{Ward Identities of $W_{\infty}$ symmetry}
\setcounter{equation}{0}
\indent

    There is an infinite number of BRST invariant states
at discrete momenta called the discrete states~\cite{bmp}.
Here we consider the discrete states of ghost number zero,
$B_{r,~s}$ and one, $\Psi_{r,~s}$ with the parametrization by two
negative integers $r$ and $s$, which have the Liouville
and the matter momenta $\alpha_{r,~s}=2+r+s$ and $\beta_{r,~s}=r-s$.
The states $\Psi_{r,~s}$ are the remnants of the massive string
modes, which are constructed from the OPE of the tachyon operators
with special momenta,
\bb
      V^-_{-r+1}(z,\z)V^+_{-s+1}(w,\w)
           \sim ~\fr{1}{\vert z-w \vert^2}
                R_{r,~s}(w) {\bar R}_{r,~s}(\w) ~,
\ee
where $V^{\pm}(z,\z)$ is the exponential part of the tachyon
operator (2.2) and  $R_{r,~s}(z)= (b_{-1} \Psi_{r,~s})(z)$.

    The $W_{\infty}$ symmetry currents are constructed from these
states as is shown below~\cite{w,k2,h}.
The discrete states $R_{r,~s} ~(r,~s \in {\bf Z}_-)$ form the
$W_{\infty}$ algebra. Here we normalize the fields such that
\bb
     R_{r,~s}(z) R_{r^{\pp},~s^{\pp}}(w) = \fr{1}{z-w}
       (r s^{\pp} - r^{\pp} s) R_{r+r^{\pp}+1,~s+s^{\pp}+1}(w) ~.
\ee
The ghost number zero states $B_{r,~s}$ have the ring structure
\bb
    B_{r,~s}(z) B_{r^{\pp},~s^{\pp}}(w) =
         B_{r+r^{\pp}+1,~s+s^{\pp}+1}(w) ~.
\ee
Combining $R_{r,~s}(z)$ and ${\bar B}_{r,~s}(\z)$, we can construct the
symmetry currents
\bb
     W_{r,~s}(z,\z) = R_{r,~s}(z) {\bar B}_{r,~s}(\z) ~,
\ee
which satisfy
\bb
     \pd_{\z} W_{r,~s}(z,\z)
         = \{ {\bar Q}_{BRST},[{\bar b}_{-1}, W_{r,~s}(z,\z)] \}~.
\ee
In the following we consider the Ward identities of the currents
\bb
     \fr{1}{\pi} \int d^2 z \pd_{\z} < W_{r,~s}(z,\z)
          ~{\cal O} >_g = 0 ~,
                        \qquad r,s \in {\bf Z}_- ~,
\ee
where ${\cal O}$ is a product of the normalized tachyon
operators (2.3).

   Let us calculate the operator product expansions (OPE) between the
currents and the tachyon operators. They are given by taking the
$c_M \rightarrow 1$ limit of the previous work for $c_M <1$~\cite{h}.
For $c_M <1$ theory the tachyon operator ${\hat O}_j$, which is
identified with the gravitationally dressed scaling operator, has
discrete momentum parametrized by the positive integer $j$.
Taking the limit
$p,~q \rightarrow \infty ~(p-q=\mbox{finite})$~\footnote{
Here $c_M = 1-6(p-q)^2 /pq$.}
and $j/q \rightarrow k$  such that
$\fr{1}{\pi}{\hat O}_j \rightarrow \Tp_k $, we then obtain
\bba
   &&   W_{-n,-m}(z,\z)~ \Tp_{k_1}(0,0)~ \Tp_{k_2} \cdots \Tp_{k_n}
               \nonumber \\
   && \quad  = \fr{1}{z}~ n! ~\biggl( \prod^n_{i=1} k_i \biggr)
                 ~ \Tp_{k_1 + \cdots +k_n -n +m}(0,0) ~,
\eea
where $\Tp_k (z,\z)$ is defined by replacing the integral in (2.2) with
${\bar c}(\z)c(z)$. This contribution is graphically expressed in fig.1.
Note that the OPE with the zero momentum tachyon $T_0$ vanishes, but
the OPE with the tachyon background $\mu_B T_0$ becomes finite due to the
renormalization (2.5).

  The OPE with the tachyon $\Tm_p$ is easily calculated by changing the
chirality. It is carried out by changing the sign of the field $X$ such
that $\Tp \rightarrow \Tm$ and
$W_{r,~s} \rightarrow -W_{s,~r}$~\footnote{
In detail $R_{r,~s}\rightarrow -R_{s,~r}$ and
$B_{r,~s} \rightarrow B_{s,~r}$. These changes preserve the OPE's (3.2) and
(3.3) respectively.}.
We then get
\bba
  &&  W_{-n,-m}(z,\z)~ \Tm_{p_1}(0,0)~ \Tm_{p_2} \cdots \Tm_{p_m}
                  \nonumber \\
  && \quad  = -\fr{1}{z} ~m! ~\biggl( \prod^m_{j=1} p_j \biggr)
                 ~ \Tm_{p_1 + \cdots +p_m +n -m}(0,0) ~.
\eea
The OPE's (3.7) and (3.8) have already been computed in~\cite{k2}.

   The derivative $\pd_{\z}$ in (3.6) picks up the OPE singularities
(3.7) and (3.8), which give the linear terms of the Ward identities.
In addition we get the correlator
\bb
      <\fr{1}{\pi} \int d^2 z \{ {\bar Q}_{BRST},
                    [{\bar b}_{-1}, W_{r,~s}(z,\z)] \} ~{\cal O} >_g ~.
\ee
This correlator does not vanish, which gives the anomalous contributions
from the boundary of moduli space as in the case of $c_M <1$ theory.
The boundary is described by using the propagator in the form
\bb
    D = \fr{1}{\pi} \int_{\e^{-\tau} \leq |z| \leq 1}
          \fr{d^2 z}{|z|^2}z^{L_0}\z^{{\bar L}_0}
      = 2 \biggl( \fr{1}{H} - \fr{1}{H} \e^{-\tau H} \biggr) ~,
\ee
where $H=L_0 +{\bar L}_0 $ and $L_0$ is the zero mode of the Virasoro
generator including the ghost part: $L_0 = L^L_0 +L^M_0 +L^G_0$.
The last term, at $\tau \rightarrow \infty$, stands for the boundary of
the moduli space.
The anomalous contributions are also calculated by taking the
limit $p,~q \rightarrow \infty ~(p-q=\mbox{finite})$
and then by replacing the summation of the discrete tachyon momentum
with  the integral of the continuous one such as, for example,
$\fr{\pi}{q} \sum^{nq-1}_{j=1} \rightarrow \int^n_0 dl $
in ref.~\cite{h}.

   There are many types of anomalous contributions. The boundary
configuration that gives the product of two amplitudes is shown in
fig.2. At $\tau \rightarrow \infty$ it gives the following contribution
for $(r,s)=(-n,-m)$:
\bb
     \lambda ~n! ~\biggl( \prod^{n-1}_{i=1} k_i \biggr) ~
          \int_0^{\sum^{n-1}_{i=1}k_i-n+m} dl
           < {\cal O}^{\pp}_1 ~\Tp_{\sum^{n-1}_{i=1}k_i-n+m-l} >_{g_1}
           < \Tp_{l} ~{\cal O}^{\pp}_2 >_{g_2} ~,
\ee
where $g_1 +g_2 =g$. The primes on ${\cal O}_1$ and ${\cal O}_2$ stand
for the exclusion of the operators $\Tp_{k_i} ~(i=1,\cdots,n-1)$, where
$\Tp_{k_i}$'s are the operators in ${\cal O}$ or the tachyon
background $-\mu_B T_0$.  The partition of the set ${\cal O}$
into ${\cal O}^{\pp}_1$ and ${\cal O}^{\pp}_2$ is done
in such a way that there is no overcounting.
The partition of genus $g$ into $g_1$ and $g_2$ is also
done.

  There is a variant of the contribution (3.11). The boundary
configuration that the surfaces of $g_1$ and $g_2$ are connected by a
handle (see fig.3) gives the following contribution:
\bb
     \lambda ~n! ~\biggl( \prod^{n-1}_{i=1} k_i \biggr) ~
          \int_0^{\sum^{n-1}_{i=1}k_i-n+m} dl
            \fr{1}{2!}< \Tp_{\sum^{n-1}_{i=1}k_i-n+m-l}\Tp_{l}
                    ~{\cal O}^{\pp} >_{g-1} ~,
\ee
where the factor $1/2!$ corrects the double counting coming from the
interchange of $\Tp_{\sum^{n-1}_{i=1}k_i-n+m-l}$ and $\Tp_{l}$.

  The boundary contribution with the triple product of amplitudes is also
calculated by using the result in ref.\cite{h} in the form
\bba
   &&  \lambda^2 ~n! ~\biggl( \prod^{n-2}_{i=1} k_i \biggr) ~
          \int_0^{\sum^{n-2}_{i=1}k_i-n+m} dl
          \int_0^{\sum^{n-2}_{i=1}k_i-n+m-l} dl^{\pp}
                   \nonumber  \\
   && \qquad  \times
          < {\cal O}^{\pp}_1
              ~\Tp_{\sum^{n-2}_{i=1}k_i-n+m-l-l^{\pp}} >_{g_1}
          < \Tp_{l} ~{\cal O}^{\pp}_2 >_{g_2}
          <\Tp_{l^{\pp}} ~{\cal O}^{\pp}_3 >_{g_3} ~,
\eea
where $g_1 +g_2 +g_3 =g$. There are some variants of the contributions
(3.13). For instance, from the configuration
that the surfaces of $g_1$ and $g_2$ are connected by a handle, we get the
contribution given by carrying out the following replacement for the
formula (3.13):
\bba
     &&  < {\cal O}^{\pp}_1
           ~\Tp_{\sum^{n-2}_{i=1}k_i-n+m-l-l^{\pp}} >_{g_1}
         < \Tp_{l} ~{\cal O}^{\pp}_2 >_{g_2}
                     \nonumber \\
     && \qquad\qquad    \rightarrow
          \fr{1}{2!} < \Tp_{\sum^{n-2}_{i=1}k_i-n+m-l-l^{\pp}} \Tp_{l}
                 ~{\cal O}^{\pp}_{1+2} >_{g_1 +g_2 -1}~.
\eea
In the case that the three surfaces are connected by handles with each other,
we then get the contribution given by replacing the triple product term with
\bb
          \fr{1}{3!} < \Tp_{\sum^{n-2}_{i=1}k_i-n+m-l-l^{\pp}}
                       \Tp_{l}\Tp_{l^{\pp}}
                 ~{\cal O}^{\pp} >_{g -2}~.
\ee

  In general the  contributions are expressed as follows (see fig.4):
\bba
   &&  \lambda^{a-1} ~n! ~\biggl( \prod^{n+1-a}_{i=1} k_i \biggr)
              \int \prod^a_{i=1} dl_i \theta(l_i)
                   ~\delta \biggl( \sum^a_{i=1}l_i
                        -\sum^{n+1-a}_{i=1} k_i +n-m \biggr)
                  \nonumber  \\
   && \qquad\qquad  \times   < \Tp_{l_1} ~{\cal O}^{\pp}_1  >_{g_1}
             < \Tp_{l_2} ~{\cal O}^{\pp}_2 >_{g_2}
                   \cdots <\Tp_{l_a} ~{\cal O}^{\pp}_a >_{g_a} ~,
\eea
where $\sum^a_{i=1}g_i =g$ and $ a=1, \cdots ,n+1$.
$\theta$ is the step function. The $a=1$ formula is nothing
but the contribution of the OPE (3.7). In addition, as discussed in the
cases of $a=2$ and $3$, there are many variants of this expression,
which come from the boundary configurations that some of the surfaces are
connected by handles.

  For the negative chirality we get
\bba
   &&  -\lambda^{b-1} ~m! ~\biggl( \prod^{m+1-b}_{j=1}p_j \biggr)
                  \int \prod^b_{i=1} dl_i \theta (l_i)
                    ~\delta \biggl( \sum^b_{i=1} l_i
                           -\sum^{m+1-b}_{j=1}p_j -n+m \biggr)
                  \nonumber  \\
   && \qquad\qquad  \times   < \Tm_{l_1} ~{\cal O}^{\pp}_1 >_{g_1}
         < \Tm_{l_2} ~{\cal O}^{\pp}_2 >_{g_2} \cdots
            <\Tm_{l_b} ~{\cal O}^{\pp}_b >_{g_b} ~,
\eea
where $\sum^b_{i=1}g_i =g$ and $ b=1, \cdots , m+1$. $b=1$ corresponds
to the contribution of the OPE (3.8). And also there are many variants
of these contributions as mentioned above.

\section{Recursion Relations}
\setcounter{equation}{0}
\indent

  In this section we write out several recursion relations and
discuss the structures of them.  We consider the recursion
relations for sphere amplitudes at first and then discuss the case of
higher genus ones. For the higher genus cases we need to add the extra
boundaries not discussed in Sect.3.

\subsection{Sphere amplitudes}
\indent

Let us first consider the $W_{-n,-1}$ identity of type
\bb
     \fr{1}{\pi} \int {\bar \pd} < W_{-n,-1}
                  ~\Tp_{k_1} ~\prod^M_{j=1} \Tm_{p_j} >_0 = 0  ~.
\ee
For the first few values of $M$, these equations become as follows.
$M=1$ gives the linear identity
\bba
   &&   -x < \Tp_{k_1} \Tm_{p_1} \Tm_{n-1} >_0
        - p_1 < \Tp_{k_1} \Tm_{p_1+n-1} >_0
               \nonumber \\
   &&\qquad\qquad\qquad\qquad
          + n x^{n-1} k_1 < \Tp_{k_1 -n+1} \Tm_{p_1} >_0 = 0 ~.
\eea
where $x=-\mu$. The first and the second terms of the l.h.s. come from
the OPE (3.8) with the background $-\mu_B T_0 ~(=\lim_{p \rightarrow 0}
(x/p)~\Tm_p)$ and the tachyon
$\Tm_{p_1}$ respectively. The third term comes from the OPE (3.7)
with $\Tp_{k_1}$ and $n-1$ $-\mu_B T_0$'s $(=\lim_{k \rightarrow 0}
(x/ k)~\Tp_k)$, where
$n = n!/(n-1)!$ and $(n-1)!$ denotes the permutation of
$-\mu_B T_0$'s.
For $M=2$ the identity becomes
\bba
    && - x < \Tp_{k_1} \Tm_{p_1} \Tm_{p_2} \Tm_{n-1} >_0
       -p_1 < \Tp_{k_1} \Tm_{p_1 +n-1} \Tm_{p_2} >_0
               \nonumber  \\
    && -p_2 < \Tp_{k_1} \Tm_{p_1} \Tm_{p_2 +n-1} >_0
       + n x^{n-1} k_1 < \Tp_{k_1 -n+1} \Tm_{p_1} \Tm_{p_2} >_0
                          \\
    && + \lambda n(n-1) x^{n-2} k_1 \int^{k_1 -n+1}_0 dl
            < \Tm_{p_1} \Tp_{k_1 -n+1-l}>_0 < \Tp_l \Tm_{p_2} >_0 = 0 ~.
                  \nonumber
\eea
The last non-linear term comes from the anomalous contribution (3.11).
For $M=3$ a triple term appears in the expression,
\bba
    && - x < \Tp_{k_1} \Tm_{p_1} \Tm_{p_2} \Tm_{p_3} \Tm_{n-1} >_0
       -p_1 < \Tp_{k_1} \Tm_{p_1 +n-1} \Tm_{p_2} \Tm_{p_3} >_0
               \nonumber  \\
    &&  -p_2 < \Tp_{k_1} \Tm_{p_1} \Tm_{p_2 +n-1} \Tm_{p_3} >_0
        -p_3 < \Tp_{k_1} \Tm_{p_1} \Tm_{p_2} \Tm_{p_3 +n-1} >_0
                \nonumber  \\
    && + n x^{n-1} k_1 < \Tp_{k_1 -n+1} \Tm_{p_1}
                      \Tm_{p_2} \Tm_{p_3} >_0
                         \\
    && + \lambda n(n-1) x^{n-2} k_1 \int^{k_1 -n+1}_0 dl
           \biggl[  < \Tm_{p_1} \Tm_{p_2} \Tp_{k_1 -n+1-l}>_0
                             < \Tp_l \Tm_{p_3} >_0
                  \nonumber  \\
    && \qquad\qquad\qquad\qquad\quad
               + < \Tm_{p_1} \Tm_{p_3} \Tp_{k_1 -n+1-l}>_0
                             < \Tp_l \Tm_{p_2} >_0
                    \nonumber \\
    && \qquad\qquad\qquad\qquad\qquad\quad
               + < \Tm_{p_2} \Tm_{p_3} \Tp_{k_1 -n+1-l}>_0
                             < \Tp_l \Tm_{p_1} >_0    \biggr]
                    \nonumber  \\
    &&  + \lambda^2 n(n-1)(n-2) x^{n-3} k_1
              \int_0^{k_1 -n+1} dl \int_0^{k_1 -n+1-l} dl^{\pp}
                      \nonumber   \\
    &&  \qquad\qquad\qquad \times
                < \Tm_{p_1} \Tp_{k_1 -n+1-l -l^{\pp}} >_0
                < \Tp_{l} \Tm_{p_2} >_0
                < \Tp_{l^{\pp}} \Tm_{p_3} >_0 = 0 ~.
                   \nonumber
\eea
In general an $M$-ple term appears for $M \leq n$, but for $M > n$
at most an $n$-ple term does. Note that in this case the $(n+1)$-ple
term vanishes by the Seiberg condition.

   We also consider the Ward identities of type
\bb
     \fr{1}{\pi} \int {\bar \pd} < W_{-n-1,-2}
                  ~\Tp_{k_1} ~\prod^M_{j=1} \Tm_{p_j} >_0 = 0  ~.
\ee
{}From the $M=1$ equation we get
\bba
   &&   -x^2 < \Tp_{k_1} \Tm_{p_1} \Tm_{n-1} >_0
        - 2! x p_1 < \Tp_{k_1} \Tm_{p_1+n-1} >_0
               \nonumber \\
   &&\qquad\qquad\qquad\qquad
          + (n+1) x^n k_1 < \Tp_{k_1 -n+1} \Tm_{p_1} >_0 = 0 ~.
\eea
After removing the amplitude $< \Tp_{k_1} \Tm_{p_1} \Tm_{n-1} >_0$ by
using the eqs.(4.2) and (4.6), we obtain
\bb
     -p_1 <\Tp_{k_1} \Tm_{p_1 +n-1} >_0
       + x^{n-1} k_1 <\Tp_{k_1 -n+1}\Tm_{p_1} >_0 = 0~.
\ee
Solving this equation we get
$<\Tp_{q_1}\Tm_{q_2}>_0 = (A / \lambda)\delta (q_1 -q_2)q_1 x^{q_1+c}$,
where $A$ and $c$ are constants. We also, using eq.(4.2) or
(4.6), get
$<\Tp_{q_1}\Tm_{q_2}\Tm_{q_3}>_0 = (A / \lambda)\delta (q_1 -q_2 -q_3)
q_1 q_2 q_3 x^{q_1 -1+c}$. To determine the constants $A$ and $c$ we
further consider the $M=2$ identity of (4.5),
\bba
    && - x^2 < \Tp_{k_1} \Tm_{p_1} \Tm_{p_2} \Tm_{n-1} >_0
       -2! p_1 p_2 <\Tp_{k_1}\Tm_{p_1 +p_2 +n-1} >_0
               \nonumber  \\
    &&  -2! x p_1 < \Tp_{k_1} \Tm_{p_1 +n-1} \Tm_{p_2} >_0
        -2! x p_2 < \Tp_{k_1} \Tm_{p_1} \Tm_{p_2 +n-1} >_0
               \nonumber     \\
    &&   + (n+1) x^n k_1 < \Tp_{k_1 -n+1} \Tm_{p_1} \Tm_{p_2} >_0
                           \\
    &&   + \lambda (n+1)n x^{n-1} k_1 \int^{k_1 -n+1}_0 dl
            < \Tm_{p_1} \Tp_{k_1 -n+1-l}>_0 < \Tp_l \Tm_{p_2} >_0 = 0 ~.
                  \nonumber
\eea
Solving the simultaneous equation of (4.3) and (4.8), we obtain $A=1$,
$c=0$ and the $1 \rightarrow 3$ amplitude.
Substituting the $1 \rightarrow 1,2,3$ amplitudes into eq.(4.4) we get
the $1 \rightarrow 4$ amplitude.

  The $1 \rightarrow N-1$ amplitude is given in order by solving the
Ward identity (4.1) (or (4.5)),
\bb
      \S^{(0)}_{q_1 \rightarrow q_2}
           = \lambda^{-1} \delta (q_1 -q_2 )~ q_1 ~x^{q_1}
\ee
and
\bb
     \S^{(0)}_{q_1 \rightarrow q_2 ,\cdots ,q_N}
           = \lambda^{-1} \delta \biggl( q_1 - \sum^N_{j=2}q_j \biggr)
                \prod^N_{j=1} q_j  \prod^{N-3}_{t=1}(q_1 -t)
                    ~x^{q_1 -N+2}
\ee
for $N>2$. We check this formula up to $N=6$ by using the $M=4$ identity
of (4.1).

   Next we consider the Ward identities with two $\Tp$ tachyons,
\bb
          \fr{1}{\pi} \int {\bar \pd} < W_{-n,-1}
              ~\Tp_{k_1} \Tp_{k_2} \prod^M_{j=1} \Tm_{p_j} >_0 = 0 ~.
\ee
For $M=1$ we get the equation
\bba
   &&  -x < \Tp_{k_1} \Tp_{k_2}\Tm_{p_1} \Tm_{n-1} >_0
        -p_1 < \Tp_{k_1} \Tp_{k_2} \Tm_{p_1 +n-1} >_0
                    \nonumber \\
   &&  + n(n-1) x^{n-2} k_1 k_2 < \Tp_{k_1 +k_2 -n+1} \Tm_{p_1} >_0
                    \nonumber   \\
   &&  + n x^{n-1} \Bigl[ k_1 < \Tp_{k_1 -n+1} \Tp_{k_2} \Tm_{p_1} >_0
                + k_2 < \Tp_{k_1} \Tp_{k_2 -n+1} \Tm_{p_1} >_0  \Bigr]
                    \nonumber \\
   &&  -\lambda \int^{n-1}_0 dl \Bigl[
                 < \Tp_{k_1} \Tm_{p_1} \Tm_{n-1-l} >_0
                   < \Tm_l \Tp_{k_2} >_0
                         \\
   && \qquad\qquad\qquad\qquad
                 + < \Tp_{k_1}  \Tm_{n-1-l} >_0
                   < \Tm_l \Tm_{p_1}\Tp_{k_2} >_0  \Bigr] = 0 ~.
                     \nonumber
\eea
The first term is the $2 \rightarrow 2$ amplitude. The other terms are
given by the solutions (4.9-10). Thus we can get the $2 \rightarrow 2$
amplitude. To solve eq.(4.12) we divide the kinematical regions into
three cases: (I) $p_1 > k_1, ~k_2 > n-1$, (II) $n-1 > k_1, ~k_2 > p_1$,
(III) $k_1 > p_1,~n-1 > k_2$ or $k_2 > p_1,~n-1 > k_1$.
In case (I) the non-linear terms (the fifth term) vanish.
In case (II) the fourth term vanishes
due to the Seiberg condition, where note that $k_1 -n+1<0$ and
$k_2 -n+1 <0$. In  case (III)  one side in the brackets of the
fourth term vanishes and also one of the non-linear terms does.
The solution is finally summarized in the form
\bba
     \S^{(0)}_{q_1,q_2 \rightarrow q_3,q_4}
       & = &\lambda^{-1} \delta (q_1 +q_2 -q_3 -q_4)
                 \prod^4_{j=1} q_j ~\Bigl( q_{max}-1 \Bigr)
                    ~x^{q_1 +q_2 -2}
                \nonumber  \\
       & = &\lambda^{-1} \delta (q_1 +q_2 -q_3 -q_4)
                 \prod^4_{j=1} q_j
                      \\
       &  & \qquad\quad \times
                    \half ~\Bigl( \vert q_1 +q_2 \vert
                           + \vert q_1 -q_3 \vert +\vert q_1 -q_4 \vert
                      -2 \Bigr) ~ x^{q_1 +q_2 -2} ~,
                    \nonumber
\eea
where $q_{max}= max(q_j)$.

   Solving the $M=2$ equation of (4.11) we get the $2 \rightarrow 3$
amplitude for all kinematical regions,
\bb
     \S^{(0)}_{q_1,q_2 \rightarrow q_3,q_4,q_5}
        = \lambda^{-1} \delta (q_1 +q_2 -q_3 -q_4 -q_5)
                 \prod^5_{j=1} q_j ~f(q)
                    ~x^{q_1 +q_2 -3}
\ee
where
\bb
      f(q) = \left\{
      \begin{array}{lc}
         ( q_{max}-1 )( q_{max}-2 )&
                     \mbox{for} \quad (q_{max},q_{min})
                              =(q_1,q_2) ~\mbox{or}~ (q_2,q_1)
                       \\
         ( q_{max} -1 )( q_{(2)} +q_{(3)} -2 ) &
                     \mbox{otherwise}
      \end{array} \right.
\ee
where $q_{min}=min(q_j)$.  $q_{(2)}$ and $q_{(3)}$ are the second and
the third largest momenta in $\{ q_j \}$.

\subsection{Higher genus amplitudes}
\indent

    Let us first consider the Ward identities of $(r,s)=(-2,-1)$,
\bb
        \fr{1}{\pi} \int {\bar \pd} < W_{-2,-1}
                \Tp_{k_1} \prod^M_{j=1}\Tm_{p_j} >_g = 0 ~.
\ee
For $M=1$ we get the equation
\bba
     && -x < \Tp_{k_1} \Tm_{p_1} \Tm_1 >_g
        -p_1 <  \Tp_{k_1} \Tm_{p_1 +1} >_g
                \nonumber \\
     && + 2! x k_1 < \Tp_{k_1 -1} \Tm_{p_1} >_g
                 \nonumber    \\
     && -\fr{\lambda}{2} \int^1_0 dl
                 < \Tm_{1-l} \Tm_l \Tm_{p_1} \Tp_{k_1} >_{g-1}
                      \\
     && +\fr{\lambda}{2} 2! k_1 \int^{k_1 -1}_0 dl
                 < \Tp_{k_1 -1 -l} \Tp_l \Tm_{p_1} >_{g-1} = 0
                \nonumber
\eea
and for $M=2$ we obtain
\bba
     && -x < \Tp_{k_1} \Tm_{p_1} \Tm_{p_2} \Tm_1 >_g
        -p_1 <  \Tp_{k_1} \Tm_{p_1 +1} \Tm_{p_2} >_g
                \nonumber \\
     && -p_2 <  \Tp_{k_1} \Tm_{p_1} \Tm_{p_2 +1} >_g
        + 2! x k_1 < \Tp_{k_1 -1} \Tm_{p_1} \Tm_{p_2} >_g
                 \nonumber  \\
     && -\fr{\lambda}{2} \int^1_0 dl
           < \Tm_{1-l} \Tm_l \Tm_{p_1} \Tm_{p_2} \Tp_{k_1} >_{g-1}
                        \\
     && +\fr{\lambda}{2} 2! k_1 \int^{k_1 -1}_0 dl
                 < \Tp_{k_1 -1 -l} \Tp_l \Tm_{p_1}\Tm_{p_2} >_{g-1}
                      \nonumber \\
     && + \lambda 2! k_1 \sum^g_{h = 0} \int^{k_1 -1}_0 dl
                     < \Tm_{p_1} \Tp_{k_1 -1-l} >_h
                     < \Tp_l \Tm_{p_2} >_{g-h}
                = 0 ~.
               \nonumber
\eea

   We can easily check that these equations satisfy the following
higher genus amplitudes.
For the $1 \rightarrow N-1$ amplitudes at
genus one and two they are given by
\bb
      \S^{(1)}_{q_1 \rightarrow q_2, \cdots, q_N}
          = \fr{1}{24} \delta \biggl( q_1 - \sum^N_{j=2} q_j \biggr)
                ~\prod^N_{j=1} q_j  \prod^{N-1}_{t=1}(q_1 -t)
                  ~ \biggl( \sum^N_{j=2}q_j^2 -q_1 -1 \biggr)
                   ~x^{q_1 -N}
\ee
and
\bba
    &&  \S^{(2)}_{q_1 \rightarrow q_2, \cdots, q_N}
          = \fr{\lambda}{5760} \delta \biggl( q_1 - \sum^N_{j=2} q_j \biggr)
                ~\prod^N_{j=1} q_j  \prod^{N+1}_{t=1}(q_1 -t) ~x^{q_1 -N-2}
                \nonumber  \\
    && \qquad\qquad\qquad\qquad \times
            \biggl[ 3\sum^N_{j=2}q^4_j
                    +10\sum^N_{i,j=2 \atop (i<j)} q^2_i q^2_j
                    -(10 q_1 +5) \sum^N_{j=2}q^2_j
             \nonumber   \\
   && \qquad\qquad\qquad\qquad\qquad
                    +10\sum^N_{i,j=2 \atop (i<j)} q_i q_j
                    +12q_1 +7  \biggr]
\eea
for $N \geq 1$. The $2 \rightarrow 2$ amplitudes at genus one is
\bba
    &&  \S^{(1)}_{q_1,q_2 \rightarrow q_3, q_4}
          = \fr{1}{24} \delta (q_1 +q_2 -q_3 -q_4) \prod^4_{i=1}q_i
               ~(q_{max}-1)~x^{Q-4}
              \nonumber    \\
    && \qquad\qquad \times
          \Bigl[(Q^3 +6Q+1)(Q-2q_{max}-6)
          +(Q^2 +L) \bigl( q^2_{max} +13q_{max} \bigr)
               \nonumber       \\
    && \qquad\qquad\qquad\qquad
           -4Q \bigl( 2q^2_{max} -9 \bigr)
           -6L( 3q_{max}-1 ) +10 q^2_{max}  \Bigr] ~,
\eea
where $Q=q_1 +q_2 ~(=q_3 +q_4)$ and $L=q_{(2)}^2 +q_{(3)}^2 $.
$q_{(2)}$ and $q_{(3)}$ are the second and the third largest momenta.
The first several of (4.19-20) and (4.21) have been derived in the
matrix model~\cite{k}.
Thus we confirm that eqs.(4.17) and (4.18) and also $M>2$
equations of (4.16) are consistent with the matrix model results.

   However, the higher genus amplitudes do not satisfy the general
$W_{-n,-m}$ identity with $n,m >2$. This indicates that further
boundary contributions are necessary for the higher genus cases.
It is easily imagined that there are the contributions shown in
fig.5. On the basis of this figure we can speculate a generalisation
of the formula (3.16) as follows:
\bba
   &&  \lambda^{a-1+h} \int \prod^a_{i=1} dl_i \theta(l_i)
          ~D^{+(h)}_a (-l_1,\cdots,-l_a,k_1,\cdots,k_{n+1-a-2h};n,m)
                  \nonumber  \\
   && \qquad\qquad  \times   < \Tp_{l_1} ~{\cal O}^{\pp}_1  >_{g_1}
             < \Tp_{l_2} ~{\cal O}^{\pp}_2 >_{g_2}
                   \cdots <\Tp_{l_a} ~{\cal O}^{\pp}_a >_{g_a} ~,
\eea
where $\sum^a_{i=1}g_i =g-h$ and $ a=1, \cdots ,n+1-2h$.
$h$ stands for the genus of the surface $\Sigma$ in fig.5, where the
$h=2$ case is presented. $-l_i$ represents the conjugate mode of
$\Tp_{l_i}$. The $h=0$ formula is nothing but (3.16).
The $h\neq 0$ contributions exist for $g \geq h$ and $ n \geq 2h$,
where note that the $h=1$ formula would contibute in the
$W_{-2,-1}$ identities, but it vanishes due to the Seiberg condition.

  The direct calculation of the D-functions for $h \geq 1$ are very
difficult.  So we guess the forms. Recall that the discrete state
$R_{-n,-m}$ is given by the OPE of the tachyon operators
\bb
      T^-_{n+1} \times T^+_{m+1} \sim R_{-n,-m} ~.
\ee
This suggests that we could replace the operator  ${\bar \pd}W_{-n,-m}$
with the two tachyons $\Tm_{n+1}$ and $\Tp_{m+1}$.
Thus we identify the surface $\Sigma$ with the
$1 \rightarrow n+2-2h$ amplitude of genus $h$,
${\hat S}^{(h)}_{m+1,-l_1,\cdots,-l_a, k_1,\cdots,k_{n+1-a-2h}
\rightarrow n+1}$.  From this argument we guess the expressions of
the D-functions as follows:
\bba
   && D^{+(h)}_a (-l_1,\cdots,-l_a,k_1,\cdots,k_{n+1-a-2h};n,m)
                     \\
   && \quad  = \fr{\lambda^{1-h}}{(n+1)(m+1)\prod^a_{i=1}(-l_i)}
           ~{\hat S}^{(h)}_{m+1,-l_1,\cdots,-l_a,
                 k_1,\cdots,k_{n+1-a-2h} \rightarrow n+1} ~,
                 \nonumber
\eea
where the $\S$-matrix formula is applied as if $-l_i$ were positive.
The normalization is fixed by fitting the $h=0$ formula with (3.16).
Using the explicit expressions of the amplitudes (4.10), (4.19) and
(4.20), we obtain the first few of the D-functions
\bba
  && D^{+(0)}_a
        = \delta ( \sum^a_{i=1}l_i -\sum^{n+1-a}_{i=1}k_i +n-m)
            ~ n!~ \biggl( \prod^{n+1-a}_{i=1} k_i \biggr) ~,
                    \\
  && D^{+(1)}_a
        = \delta ( \sum^a_{i=1} l_i -\sum^{n-1-a}_{i=1}k_i +n-m)
            ~\fr{1}{24}~ n! ~\biggl( \prod^{n-1-a}_{i=1} k_i \biggr)
                     \nonumber \\
  && \qquad\qquad\qquad\quad \times
            \biggl( \sum^a_{i=1} l_i^2 + \sum^{n-1-a}_{i=1}k_i^2
                        +(m+1)^2 -n-2 \biggr)
\eea
and
\bba
  && D^{+(2)}_a
         = \delta ( \sum^a_{i=1}l_i -\sum^{n-3-a}_{i=1}k_i +n-m)
             ~\fr{1}{5760} ~n!
               ~\biggl( \prod^{n-3-a}_{i=1} k_i \biggr)
                   \nonumber  \\
  && \qquad\qquad \times
            \Biggl[
              3 \biggl\{ \sum^a_{i=1} l_i^4 + \sum^{n-3-a}_{i=1}k_i^4
                                           + (m+1)^4 \biggr\}
              + 10 \biggl\{ \sum^a_{i,j=1 \atop (i<j)}l_i^2 l_j^2
                      + \sum^{n-3-a}_{i,j=1 \atop (i<j)}k_i^2 k_j^2
                    \nonumber \\
  && \qquad\qquad\qquad\qquad\qquad
                      + \sum^a_{i=1}\sum^{n-3-a}_{j=1} l_i^2 k_j^2
                      + (m+1)^2 \biggl( \sum^a_{i=1}l_i^2
                      +  \sum^{n-3-a}_{i=1}k_i^2 \biggr)  \biggr\}
                    \nonumber  \\
  && \qquad\qquad\qquad\qquad
                    -\Bigl( 10(n+1)+5 \Bigr) \biggl\{ \sum^a_{i=1}l_i^2
                       +\sum^{n-3-a}_{i=1}k_i^2 +(m+1)^2 \biggr\}
                     \nonumber  \\
  && \qquad\qquad\qquad\qquad
                + 10 \biggl\{ \sum^a_{i,j=1 \atop (i<j)}l_i l_j
                      + \sum^{n-3-a}_{i,j=1 \atop (i<j)}k_i k_j
                      + \sum^a_{i=1}\sum^{n-3-a}_{j=1} (-l_i) k_j
                         \\
  && \qquad\qquad\qquad\qquad\quad
                      + (m+1) \biggl( \sum^a_{i=1}(-l_i)
                      +  \sum^{n-3-a}_{i=1}k_i \biggr) \biggr\}
                      +12(n+1) +7  \Biggr] ~.
                    \nonumber
\eea

  Similarly, for the negative chirality, we get
\bba
   &&  -\lambda^{b-1+h} \int \prod^b_{i=1} dl_i \theta (l_i)
           ~D^{-(h)}_b (-l_1, \cdots, -l_b, p_1, \cdots, p_{m+1-b-2h};n,m)
                  \nonumber  \\
   && \qquad\qquad  \times   < \Tm_{l_1} ~{\cal O}^{\pp}_1 >_{g_1}
         < \Tm_{l_2} ~{\cal O}^{\pp}_2 >_{g_2} \cdots
            <\Tm_{l_b} ~{\cal O}^{\pp}_b >_{g_b} ~,
\eea
where
\bba
   && D^{-(h)}_b (-l_1, \cdots, -l_b, p_1, \cdots, p_{m+1-b-2h};n,m)
                  \\
   && \quad = \fr{\lambda^{1-h}}{(n+1)(m+1)\prod^b_{i=1}(-l_i)}
           ~{\hat S}^{(h)}_{m+1 \rightarrow n+1,-l_1,\cdots,-l_b,
                                      p_1,\cdots,p_{m+1-b-2h}}
                 \nonumber
\eea
and $\sum^b_{i=1}g_i =g-h$ and $ b=1, \cdots , m+1-2h$.

  If the D-functions are given, we can write out all types of the
Ward identities. For instance we get
\bba
     0 &=& \fr{1}{\pi} \int d^2 z {\bar \pd}
               < W_{-n,-1}\Tp_{k_1}\Tm_{p_1} >_g
                     \nonumber \\
       &=& -p_1<\Tp_{k_1}\Tm_{p_1+n-1}>_g
                -x<\Tp_{k_1}\Tm_{p_1}\Tm_{n-1}>_g
                      \nonumber \\
       & & -\fr{\lambda}{2}\int^{n-1}_0 dl
               <\Tm_{n-1-l}\Tm_l\Tm_{p_1}\Tp_{k_1}>_{g-1}
                     \\
       & & + \sum^{g+1}_{a=1}\sum^{g-a+1}_{h=0}
              \fr{\lambda^{a-1+h}}{a!}
              \fr{(x/ \eps)^{n-2h-a}}{(n-2h-a)!}
              \int \prod^a_{i=1}dl_i \theta(l_i)
                      \nonumber  \\
       & & \qquad \times
              D^{+(h)}_a (-l_1,\cdots,-l_a,k_1,\eps,\cdots,\eps;n,1)
                <\Tm_{p_1}~\prod^a_{i=1}\Tp_{l_i} >_{g-h-a+1} ~.
                       \nonumber
\eea

    The D-functions are now given up to $h=2$ so that we can calculate
all amplitudes up to genus two. In the following we give the
procedure to derive the amplitudes more than genus 2.
To calculate the three genus amplitudes, let us consider
the Ward identities of type
\bb
    \fr{1}{\pi}\int d^2 z {\bar \pd}< W_{-3,-2} \Tp_{k_1}
                            \prod^M_{j=1}\Tm_{p_j}>_g =0 ~.
\ee
For $M=1$ this identity gives
\bba
   && -x^2 <\Tp_{k_1}\Tm_{p_1}\Tm_1 >_g
          -2x p_1 <\Tp_{k_1}\Tm_{p_1 +1}>_g
                \nonumber \\
   && -\lambda x \int^1_0 dl <\Tm_{1-l}\Tm_l
                        \Tm_{p_1}\Tp_{k_1}>_{g-1}
       -\lambda p_1 \int^{p_1 +1}_0 dl
                  <\Tm_{p_1+1-l}\Tm_l \Tp_{k_1}>_{g-1}
                 \nonumber \\
   && -\fr{\lambda^2}{3} \int^1_0 dl \int^{1-l}_0 d l^{\pp}
              <\Tm_{1-l-l^{\pp}}\Tm_l \Tm_{l^{\pp}}
                                    \Tm_{p_1}\Tp_{k_1}>_{g-2}
      -\fr{13}{12}\lambda <\Tp_{k_1}\Tm_{p_1}\Tm_1 >_{g-1}
                  \nonumber  \\
   && +3 x^2 k_1 <\Tp_{k_1-1}\Tm_{p_1}>_g
        +3\lambda x k_1 \int^{k_1-1}_0 dl
                  <\Tp_{k_1-1-l}\Tp_l\Tm_{p_1}>_{g-1}
                   \nonumber  \\
   && +\lambda^2 k_1 \int^{k_1-1}_0 dl \int^{k_1-1-l}_0 d l^{\pp}
       <\Tp_{k_1-1-l-l^{\pp}}\Tp_l \Tp_{l^{\pp}}\Tm_{p_1}>_{g-2}
                    \nonumber  \\
   && +\fr{\lambda}{4} k_1 \bigl[ (k_1-1)^2 +k_1^2 +4 \bigr]
                    <\Tp_{k_1-1}\Tm_{p_1}>_{g-1} =0 ~,
\eea
where the 6-th and the last terms of l.h.s. are the contributions
from the $D^{-(h=1)}_{a=1}$ and the $D^{+(h=1)}_{a=1}$ formulas
respectively.

  Combining the identities (4.17) and (4.32) of $g=3$ and removing the
amplitude $<\Tp_{k_1}\Tm_{p_1}\Tm_1 >_3$, we get the identity in which
the $1 \rightarrow 1$ amplitude of genus three is expressed by the
lower genus amplitudes. Substituting the explicit expressions of the
lower genus amplitudes in the identity we get
\bba
   && x p_1 < \Tp_{k_1}\Tm_{p_1 +1}>_3
         -x^2 k_1 <\Tp_{k_1 -1}\Tm_{p_1}>_3
                  \\
   && ~ =
          \fr{\lambda^2}{483840} \delta (k_1 -p_1 -1)
             ~(p_1 +1) p_1^2 (p_1 -1)(p_1 -2)(p_1 -3)(p_1 -4)
                \nonumber  \\
   && \qquad\quad \times
           \Bigl[ 18 p_1^6 -84 p_1^5 -35 p_1^4 +238 p_1^3
                   +161 p_1^2 -112 p_1 -93 \Bigr]~x^{p_1 -4}~.
             \nonumber
\eea
Solving this equation we get the $1 \rightarrow 1$ amplitude of
genus three
\bba
   &&  \S^{(3)}_{q_1 \rightarrow q_2} =
          \fr{\lambda^2}{2903040}\delta (q_1 -q_2)
            q_1^2  \prod^5_{t=1}(q_1 -t) ~x^{q_1 -6}
                    \nonumber \\
   && \qquad\qquad\quad \times
          \Bigl[ 9 q_1^6 -63 q_1^5 +42 q_1^4 +217 q_1^3
                          -205 q_1 -93  \Bigr] ~.
\eea
This agrees with the result computed in~\cite{mpr}.

  The $1 \rightarrow 2$ amplitude of genus 3 is also given by
considering the simultaneous equation of the $M=2$ identities of
(4.16) and (4.31). Removing the amplitude $<\Tp_{k_1}\Tm_{p_1}\Tm_{p_2}
\Tm_1>_3$, the $1 \rightarrow 2$ amplitude of $g=3$ is expressed
by the lower genus amplitudes and $1 \rightarrow 1$ amplitudes
of $g=3$ obtained in eq.(4.34). Recurrently we can obtain the
$1 \rightarrow N-1$ amplitude of genus three. Furthermore,
using this, we can determine the D-functions of $h=3$. Then we
can get all types of genus three amplitudes. Repeating the
calculation mentioned above we can derive amplitudes of $g >3$.

  We finally discuss the partition function of 2D string theory.
The second derivative of it with respect to the tachyon background
(cosmological constant) $x ~(= -\mu)$ is given by
\bb
     \fr{\pd^2 Z_g}{\pd x^2}
            = \biggl( \fr{\mu_B}{\mu} \biggr)^2 < T_0 ~T_0 >_g
            = \fr{1}{\eps^2}<\Tp_{\eps}\Tm_{\eps} >_g ~.
\ee
Therefore, using the resuts of the $1 \rightarrow 1$ amplitudes
of each genus, we get
\bb
    \fr{\pd^2 Z}{\pd x^2} = \lambda^{-1}\log x + \fr{1}{24} x^{-2}
       - \lambda \fr{7}{960} x^{-4}
       + \lambda^2 \fr{31}{8064}x^{-6} + \cdots ~,
\ee
where $Z = \sum_g Z_g$. The first term of r.h.s. is given by the
algebra $\fr{1}{\eps}x^{\eps}= \fr{1}{\eps}+\log x$, where the
divergent part should be regularized properly. From the second term
we get $Z_1 = -\fr{1}{24} \log x$. Note that, contrary to the $c_M <1$
case,  there is no problem with doubling of the partition functions when
we identify the matrix model result and the continuum one because
in 2D string theory we consider both chiralities, while in $c_M <1$ only
one half of the chiralities are considered.

\section{On Differences Between $c_M =1$ and $c_M <1$}
\setcounter{equation}{0}
\indent

   The scaling operator of quantum gravity coupled to $(p,q)$ minimal
matter, which is subject to the $W_q$ algebra constraints~\cite{fkn}, is
\bb
      {\hat O}_j = \fr{\Gamma(j/q)}{\Gamma(-j/q)}O_j ~,
\ee
where $j=1,2,3,\cdots ,~(j \neq q~\mbox{mod}~q)$. $O_j$ is the
tachyon operator with the discrete momentum defined in ref.~\cite{h}.
The operator ${\hat O}_{nq}, ~n \in {\bf Z}_{>0} $ decouples from the theory
since the $\Gamma$-factor in (5.1) vanishes at $j=nq$.
Recall that, as discussed in Sect.3, in the limit
$q \rightarrow \infty ~(c_M \rightarrow 1)$ and $j/q \rightarrow k$,
the scaling operator (5.1) becomes the tachyon operator with positive
chirality $\Tp_k$.

  We now clarify the structural differences between the $c_M <1$ model
satisfying the W-algebra constraints and 2D string theory.
In 2D string theory we consider both chiralities of the tachyon operators
to define the $S$-matrix, while in the $c_M <1$ model we consider only
one half of the chiralities, where it is the positive
one.\footnote{
By interchanging the roles of $p$ and $q$ we get the theory defined on
the negative chirality.}
Also in $c_M <1$ theory the operator ${\hat O}_{nq}$
decouples from the theory, but the corresponding tachyon operator $\Tp_n$
of the $c_M =1$ model no longer decouples. The factor $\Lambda(k)$ (2.3)
vanishes at the momentum $k=n$, but extra divergence
appears at $q \rightarrow \infty$ so that the $\S$-matrix becomes finite.
It means that the $S$-matrix of 2D string theory has a pole at
the momentum $k=n$ due to the external leg factor $\Lambda^{-1}(k)$.

 The pole of the $\Lambda^{-1}$-factor plays an important role
when we consider spacetime physics~\cite{np,jly},
which is related to the discrete state through
the OPE (3.1). For example,
$T^+_2 \times T^-_2 \sim \pd X {\bar \pd} X$, where the
r.h.s. is the operator giving the deformation of geometry.
In fact in ref.~\cite{np} the contribution from the $k=2$ pole gives
the gravitational scattering.
On the other hand, for $c_M <1$, the decoupling of the $j = nq$
operators indicates that the discrete states are excluded
from the theory.

  The extra boundary contributions discussed in Sect.4.2 also seem to
appear in the expressions of the W-algebra constraints
for $c_M <1$. For the positive chirality theory the Ward identity of
the current $W_{-s+1,-l-s+1}$ is identified with the
${\cal W}^{(s)}_l$ constraint~\cite{h},
where $s=2$ is the Virasoro constraint. The extra boundary contributions
appear only for $s \geq 4$ constraints, independently of $l$, because
the extra boundary for negative chirality (4.28) does not contribute
in this case~\footnote{
The necessity of the extra terms for $s \geq 4$ constraints is also
discussed in ref.~\cite{fkn2}.
}. 

\vspace{5mm}

 I am grateful to B. Bullock for careful reading of the manuscript.

\newpage

{\flushleft {\bf Figure Captions}}

\begin{itemize}
\item
Fig.1  (a) The boundary corresponding to the OPE (3.7). The incoming  and
the outgoing arrows stand for the tachyon vertex operators with $+$ and$-$
chiralities respectively. The W point is the $W_{-n,-m}$ current.
After integrating over the locations of the vertex operators
$z_2 , \cdots, z_n$, we find the $1/z$ pole.
(b) The derivative ${\bar \pd}$ picks up the singularity and then we get
the 1-b configuration. Here $K=k_1 + \cdots +k_n -n+m$.
\item
Fig.2  (a) The boundary producing the non-linear term. The cross point
stands for the operator  ${\bar \pd} W_{-n,-m}$.
(b) At $\tau \rightarrow \infty$  we get the product of two surfaces.
Here $K=k_1 + \cdots +k_{n-1}-n+m$ and $l$ is integrated from $0$ to $K$.

\item
Fig.3  This is a variant of the boundary of fig.2, where the surfaces
$\Sigma_1$ and $\Sigma_2$ are connected by a handle.
\item
Fig.4  (a) The boundary for the general formula (3.16).
(b) The configuration at $\tau \rightarrow \infty$.
\item
Fig.5  The extra boundary contribution for the case of $h=2$.
The surface $\Sigma$ has two handles.
\end{itemize}

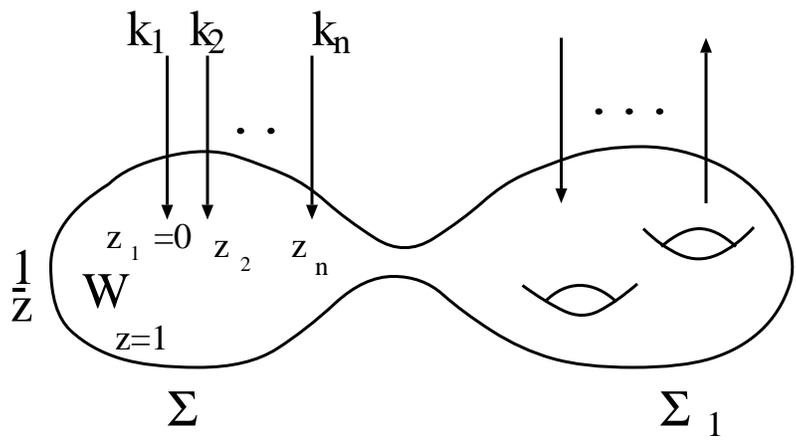
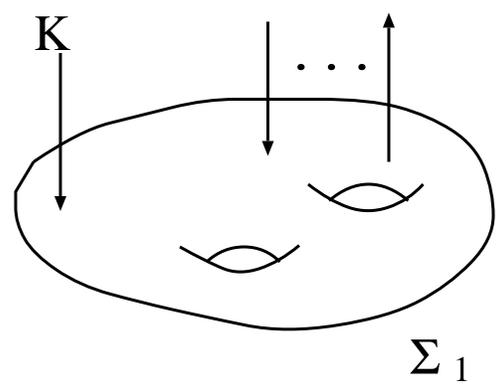
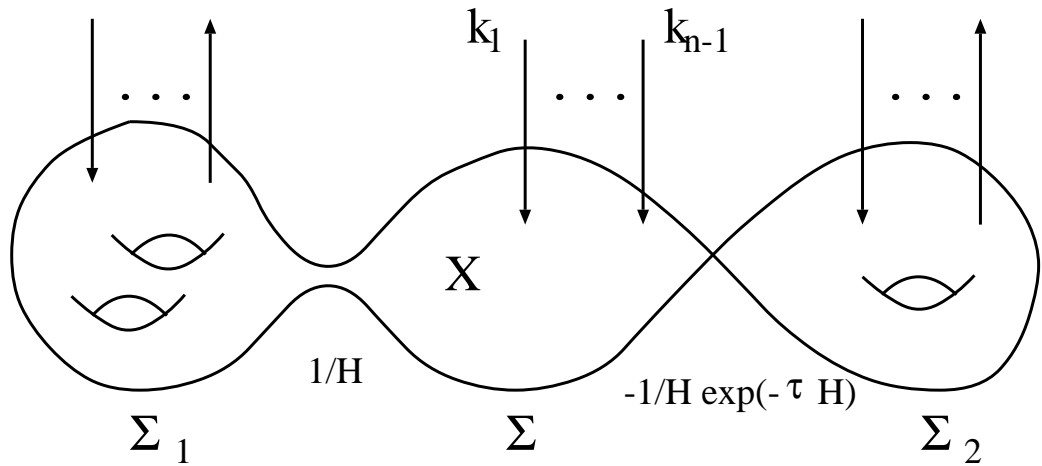
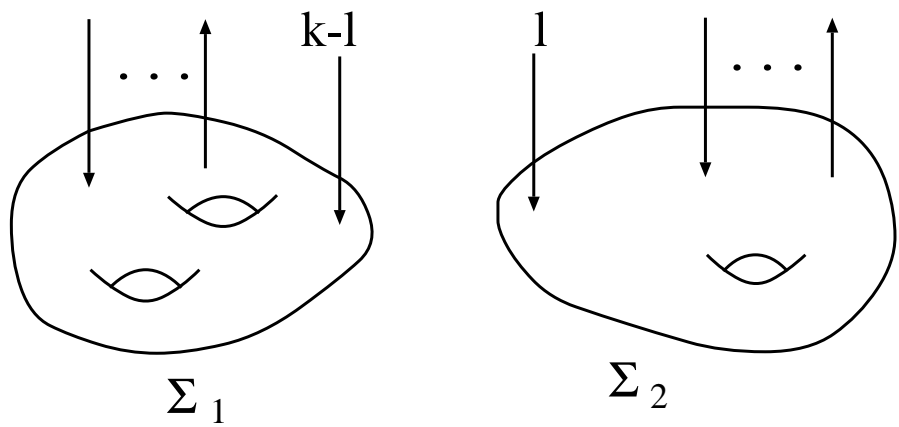
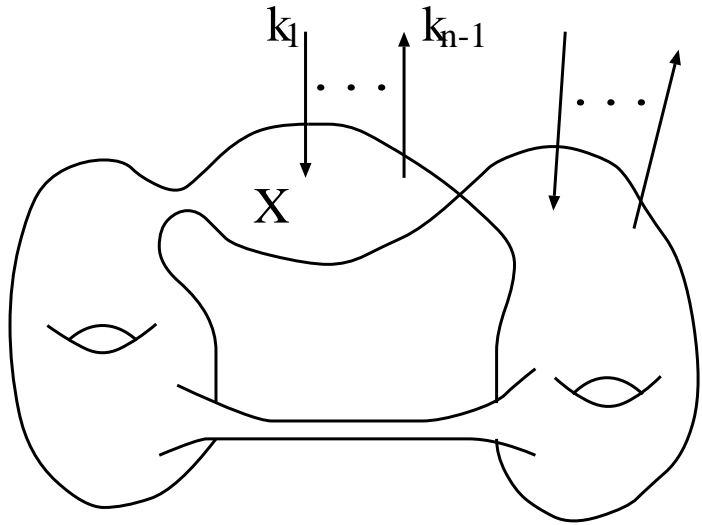
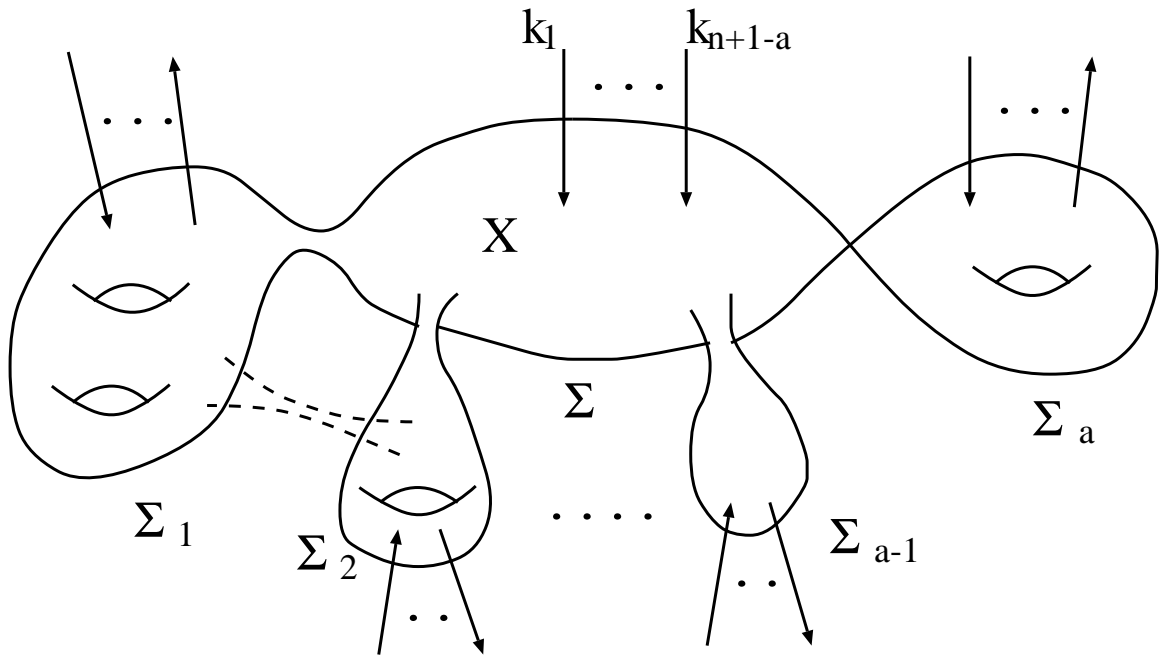
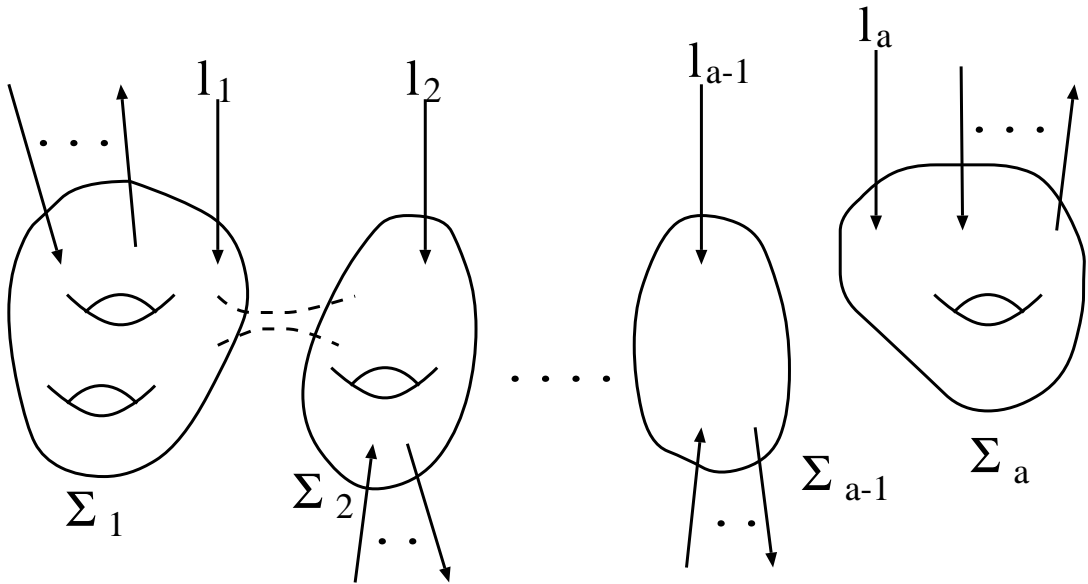
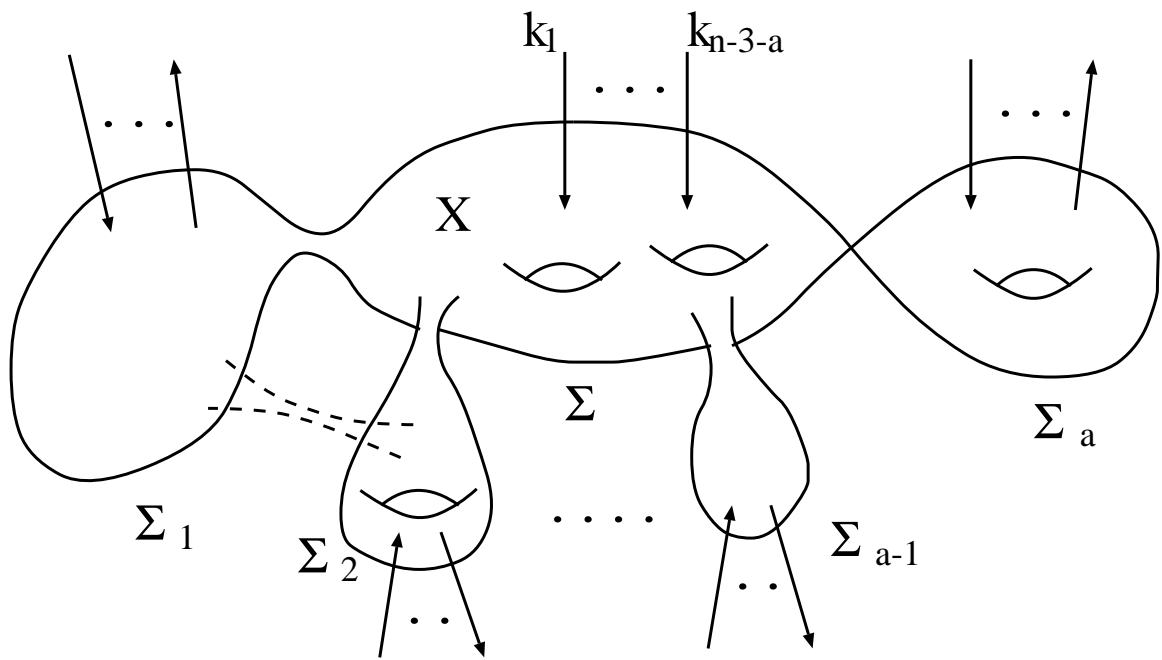
\end{document}